\documentclass[prl,reprint,twocolumn,superscriptaddress,showpacs,floatfix]{revtex4-1}

\usepackage{mathrsfs}
\usepackage{txfonts}
\usepackage{amssymb}
\usepackage{graphicx}
\usepackage{hyperref}
\usepackage{ulem}
 \usepackage{overpic}
 \usepackage{psfrag}
\usepackage{tabularx}
\usepackage{array}
\usepackage{placeins}
\newcommand{\PreserveBackslash}[1]{\let\temp=\\#1\let\\=\temp}

\makeatletter

\begin{document}

\title{Duality and ground-state phase diagram for the quantum XYZ model with arbitrary spin $s$ in one spatial dimension}

\author{Qian-Qian Shi}
\affiliation{Centre for Modern Physics, Chongqing University, Chongqing 400044, The People's Republic of China}
\author{Sheng-Hao Li}
\affiliation{Centre for Modern Physics, Chongqing University, Chongqing 400044, The People's Republic of China}
\author{Huan-Qiang Zhou}
 \email[E-mail address: ]{hqzhou@cqu.edu.cn}
\affiliation{Centre for Modern Physics, Chongqing University, Chongqing 400044, The People's Republic of China}
 \altaffiliation{ }

\begin{abstract}
Five duality transformations are unveiled for the quantum XYZ model with arbitrary spin $s$ in one spatial dimension.  The presence of these duality transformations, together with an extra Hamiltonian symmetry, drastically reduce the entire ground-state phase diagram to two {\it finite} regimes - the principal regimes, with all the other ten regimes dual to them. Combining with the determination of critical points from the conventional order parameter  approach and/or the fidelity approach to quantum phase transitions, we are able to map out the ground-state phase diagram for the quantum XYZ model with arbitrary spin $s$.  This is explicitly demonstrated for $s=1/2,1,3/2$ and 2.  As it turns out, all the critical points, with central charge $c=1$, are self-dual under a respective duality transformation for half-integer as well as integer spin $s$.  However, in the latter case, the presence of the so-called symmetry protected topological phase, i.e., the Haldane phase, results in extra lines of critical points with central charge $c=1/2$, which is not self-dual under any duality transformation.
\end{abstract}
\maketitle

{\it Introduction.}- Quantum duality is a fundamental concept that offers a powerful means to investigate critical phenomena in quantum many-body systems~\cite{ortiz}.
A prototypical example to illustrate the importance of duality is the  quantum transverse field Ising chain, an adaptation from the Kramers-Wannier duality in the two-dimensional classical Ising model~\cite{kw}.
An important lesson learned from this example is that a critical point may be identified as  a self-dual point, which is left intact under the duality transformation.  That is, under some physically sensible assumptions, the duality transformation not only establishes a  connection between two distinct phases, but also offers a practical means to locate a critical point.  However,
it remains unclear whether or not quantum duality is ubiquitous in quantum many-body systems. In addition, a relevant intriguing question is to clarify the connection between a self-dual point and a critical point.

We aim to address these two related issues with an illustrative example - the quantum XYZ model with arbitrary spin $s$, both integer and half-integer. Two special cases have been widely investigated: one is the spin-1/2 XYZ model, and the other is the $SU(2)$ Heisenberg model with arbitrary spin $s$.  Historically, the spin-1/2 XYZ model is a fundamental model in statistical physics, mainly due to the fact that the model is exactly solvable, as shown by Baxter~\cite{baxterbook} from  its equivalence to the classical two-dimensional eight-vertex model.  However, the exact solvability is lost for  the quantum XYZ model if $s$ becomes larger than 1/2.  Meanwhile, the Heisenberg model with arbitrary spin $s$  may be mapped to the nonlinear $\sigma$ model with a topological term~\cite{Haldane}. As it turns out,  the model is gapless for half-integer spin $s$ and gapped for integer spin $s$.  Generically, only a few results are available in the literature, such as the quantum XXZ model with spin $s=1$ or $s=2$~\cite{dmrgxxz,dmrgxxzspin2}.    Therefore, it is highly desirable to map out the ground-state phase diagram for the quantum XYZ model with arbitrary spin $s$.  This is achieved here by combining quantum duality with the conventional order parameter approach and/or the
fidelity approach to quantum phase transitions~\cite{zhou}.

Specifically,  five different duality transformations are unveiled for the quantum XYZ model with arbitrary spin $s$.
As it turns out, it is the presence of these duality transformations, together with an extra Hamiltonian symmetry, that drastically reduce the entire ground-state phase diagram into two {\it finite} regimes - the so-called principal regimes, with all the other ten regimes dual to them, respectively. In other words, we {\it only} need to focus on the principal regimes, in order to map out the ground-state phase diagram.   This can be done through, e.g., numerical simulations by means of the density matrix renormalization group (DMRG)~\cite{white} and the infinite time-evolving block decimation (iTEBD)~\cite{vidal}.  In this work, the iTEBD algorithm is exploited to simulate the quantum  XYZ model with $s=1/2, 1, 3/2$ and 2.  It is found that  all the critical points, with central charge $c=1$, are self-dual under a respective duality transformation for half-integer as well as integer spin $s$.  However, in the latter case, the presence of the so-called symmetry protected topological (Haldane) phase, results in extra lines of critical points with central charge $c=1/2$. These are not self-dual under any duality transformation.

{\it Quantum XYZ model.}\label{xyzmodeldual}- The Hamiltonian for the quantum XYZ model with arbitrary spin $s$ in one spatial dimension takes the form
\begin{equation}
H(\Delta,\gamma)=\sum_{i}{(\frac{1+\gamma}{2}S_i^x S_{i+1}^x+\frac{1-\gamma}{2}S_i^yS_{i+1}^y+\frac{\Delta}{2} S_i^zS_{i+1}^z)}, \label{xyzham}
\end{equation}
where $S_i^{\beta}$, with $\beta=x,y,z$, are the spin matrices for spin $s$ at site $i$, and $\gamma$ and $\Delta$ are the coupling parameters describing the anisotropic  interactions.

{\it Duality transformations for the quantum XYZ model.}- Quantum duality is a local or nonlocal nontrivial unitary transformation $U$, which leaves the form of the local Hamiltonian density intact.
Mathematically, for a Hamiltonian $H(\Delta,\gamma)$, with $\Delta$ and $\gamma$ being control parameters, $H(\Delta',\gamma')$ is dual to $H(\Delta,\gamma)$, if there is a unitary transformation $U$ such that $H(\Delta,\gamma) = k(\Delta,\gamma) U H(\Delta',\gamma')U^{\dagger}$, with $\Delta'$ and $\gamma'$ in turn being functions of $\Delta$ and $\gamma$ and $k(\Delta,\gamma)$ being positive.

For the quantum XYZ model with arbitrary spin $s$, there are five distinct duality transformations:\\
(0) The Hamiltonian $H(\Delta, \gamma)$ for $\gamma>1$ is dual to the Hamiltonian $H(\Delta', \gamma')$ for $0<\gamma<1$ under a local unitary transformation $U_0$:
$S_{2i}^x\rightarrow S_{2i}^x$, $S_{2i}^y\rightarrow S_{2i}^y$, $S_{2i}^z\rightarrow S_{2i}^z$, $S_{2i+1}^x\rightarrow S_{2i+1}^x$, $S_{2i+1}^y\rightarrow-S_{2i+1}^y$ and $S_{2i+1}^z\rightarrow-S_{2i+1}^z$: $H(\Delta, \gamma)=k(\Delta,\gamma) U_0 H(\Delta', \gamma') U_0^\dagger$,
with $\Delta'=-\Delta/\gamma$, $\gamma'=1/\gamma$, and $k(\Delta,\gamma) =\gamma$.
The Hamiltonian is self-dual if $\Delta =0$ and $\gamma=\pm 1$.

(1) Under a local unitary transformation $U_1$: $S_{i}^x\rightarrow-S_{i}^x$,
$S_{i}^y\rightarrow S_{i}^z$, $S_{i}^z\rightarrow S_{i}^y$, we have
$H(\Delta, \gamma)=k(\Delta,\gamma) U_1 H(\Delta', \gamma') U_1^\dagger$, with
$\Delta'=(2-2\gamma)/(1+\Delta+\gamma)$, $\gamma'=(1-\Delta+\gamma)/(1+\Delta+\gamma)$, and    $k(\Delta,\gamma)=(1+\Delta+\gamma)/2$.
The Hamiltonian on the line $\gamma=1-\Delta$ is self-dual.

(2) Under a local unitary transformation $U_2$:  $S_{2i}^x\rightarrow-S_{2i}^x$,  $S_{2i}^y\rightarrow-S_{2i}^z$, $S_{2i}^z\rightarrow-S_{2i}^y$, $S_{2i+1}^x\rightarrow-S_{2i+1}^x$, $S_{2i+1}^y\rightarrow S_{2i+1}^z$ and $S_{2i+1}^z\rightarrow S_{2i+1}^y$,
we have $H(\Delta, \gamma)=k(\Delta,\gamma) U_2 H(\Delta', \gamma') U_2^\dagger$, with $\Delta'=(-2+2\gamma)/(1-\Delta+\gamma)$,
$\gamma'=(1+\Delta+\gamma)/(1-\Delta+\gamma)$, and  $k(\Delta,\gamma)=(1-\Delta+\gamma)/2$.
The Hamiltonian on the line $\gamma=1+\Delta$ is self-dual.

(3) Under a local unitary transformation $U_3$: $S_{i}^x\rightarrow S_{i}^z$, $S_{i}^y\rightarrow-S_{i}^y$, $S_{i}^z\rightarrow S_{i}^x$,
we have
$H(\Delta, \gamma)=k(\Delta,\gamma) U_3 H(\Delta', \gamma') U_3^\dagger$, with $\Delta'=(2+2\gamma)/(1+\Delta-\gamma)$, $\gamma'=(-1+\Delta+\gamma)/(1+\Delta-\gamma)$, and  $k(\Delta,\gamma)=(1+\Delta-\gamma)/2$. The Hamiltonian on the line $\gamma=-1+\Delta$ is self-dual.

(4) Under a local unitary transformation $U_4$: $S_{2i}^x\rightarrow-S_{2i}^z$,  $S_{2i}^y\rightarrow-S_{2i}^y$, $S_{2i}^z\rightarrow-S_{2i}^x$, $S_{2i+1}^x\rightarrow S_{2i+1}^z$, $S_{2i+1}^y\rightarrow-S_{2i+1}^y$ and $S_{2i+1}^z\rightarrow S_{2i+1}^x$,
we have $H(\Delta, \gamma)=k(\Delta,\gamma) U_4 H(\Delta', \gamma') U_4^\dagger$, with $\Delta'=(2+2\gamma)/(-1+\Delta+\gamma)$,
$\gamma'=(-1-\Delta+\gamma)/(1-\Delta-\gamma)$, and  $k(\Delta,\gamma)=(1-\Delta-\gamma)/2$.
The Hamiltonian on the line $\gamma=-1-\Delta$ is self-dual.

\begin{figure}
     \includegraphics[angle=0,totalheight=5cm]{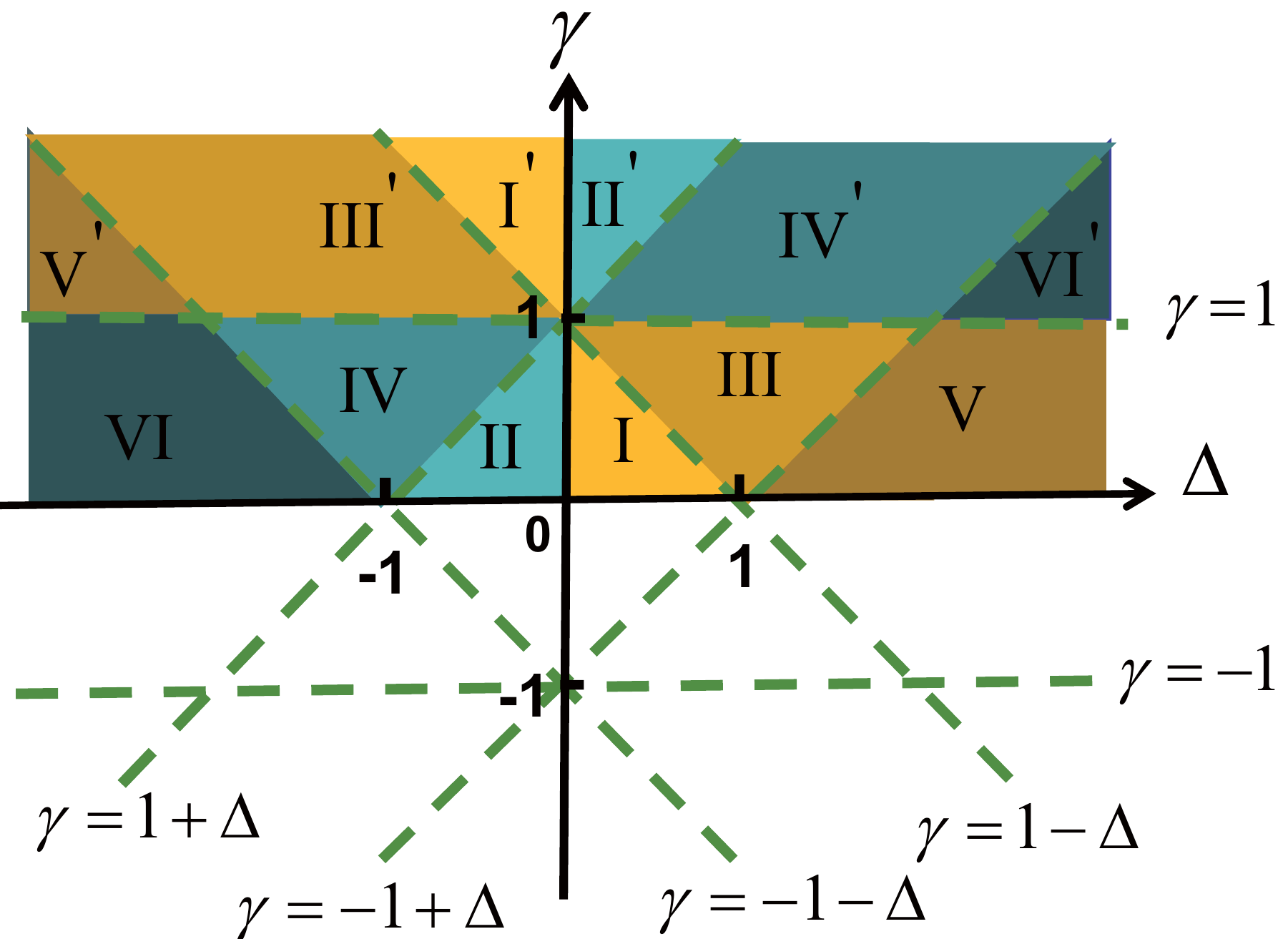}
       \caption{ Twelve regimes generated from the five duality transformations for the quantum XYZ model with arbitrary spin $s$ for $\gamma>0$. Here, the regimes $\rm{I}$, $\rm{III}$, $\rm{V}$, $\rm{I'}$, $\rm{III'}$ and $\rm{V'}$  are dual to each other.  and the regimes $\rm{II}$, $\rm{IV}$, $\rm{VI}$, $\rm{II'}$, $\rm{IV'}$ and $\rm{VI'}$ are dual to each other.  We choose the regimes $\rm I$ and $\rm {II}$ as the principal regimes. }\label{twelveregimes}
\end{figure}

Given that the Hamiltonian is symmetric with respect to  the mapping $\gamma \leftrightarrow-\gamma$: $S^x_i \leftrightarrow S^y_i$ and $S^z_i \rightarrow -S^z_i$, we may restrict ourselves to the region $\gamma \geq 0$ in the parameter space. This symmetry may be regarded as a duality transformation with $k(\Delta,\gamma)=1$, under which the line $\gamma=0$ is self-dual. Then,
as a consequence of the five distinct duality transformations, the whole region is divided, via five lines described by $\gamma=1$ and $\gamma=\pm1\pm\Delta$, into twelve different regimes, as shown in Fig.~\ref{twelveregimes}.
In addition, these twelve regimes are separated into two groups, with six regimes in each group dual to each other.  Therefore, we {\it only} need to consider two of these twelve regimes, which represent the physics underlying the quantum XYZ model with arbitrary spin $s$. For our purpose, it is convenient to choose one {\it finite} regime from each group, which is defined as a principal regime. Here and hereafter,  we choose the regimes $\rm I$ and $\rm {II}$ as the principal regimes.

{\it The iTEBD algorithm: numerical simulations.}- The iTEBD algorithm~\cite{vidal} is exploited to simulate the quantum XYZ model in the principal regimes $\rm I$ and $\rm {II}$.  The algorithm generates a ground-state wave function in a matrix product state representation on an infinite-size chain, and provides an efficient means to evaluate various physical observables.

In order to locate critical points in the principal regimes, we adopt the strategy to evaluate the order parameters for symmetry-breaking ordered phases and the string-order parameter for the Haldane phase, as explicitly presented below. The same goal may also be achieved in the context of the fidelity approach to quantum phase transitions~\cite{zhou}. Once this is done,  we are able to map out the entire ground-state phase diagram by resorting to the duality transformations.

For a given critical point, it is necessary to determine the universality class, to which it belongs. To accomplish this,  we perform a finite-entanglement scaling analysis~\cite{centralchargescaling} to extract central charge $c$ at a given critical point. Specifically, the von Neumann entropy $S$  scales with the bond dimension $\chi$ at a critical point, which is governed by a universal  pre-factor involving central charge $c$~\cite{centralchargescaling}:
\begin{equation}
S(\chi) = c \frac{\kappa \log_2{\chi}}{6}+a.
\end{equation}
Here,  $\kappa$  follows from the scaling of the correlation length with the bond dimension $\chi$: $\xi=b\chi^{\kappa}$, and $a$ and $b$ are some constants.

The quantum XYZ model also features factorized states. Although this fact has been known rigorously, we mention that a powerful numerical scheme, which is based on the geometric entanglement, is available to identify any possible factorized states~\cite{ge}.

\begin{figure}
       \includegraphics[angle=0,width=0.45\textwidth]{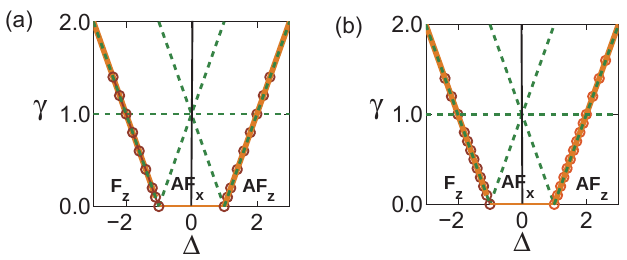}
  \caption{ Ground-state phase diagram for the quantum XYZ model with half-integer spin $s$:  (a) $s=1/2$ and (b) $s=3/2$. Here, the solid lines, i.e.,  $\gamma=0~(-1<\Delta\leq 1)$, $\gamma=-1-\Delta~(\Delta<-1)$ and $\gamma =  - 1 +\Delta ~(\Delta \geq 1)$,
 denote the phase boundaries between distinct symmetry-breaking ordered phases, labeled as $\rm {AF}_x$, $\rm {AF}_z$ and $\rm {F}_z$.  The self-dual line, $\gamma=1+\Delta$, with $\Delta>-1$, describes the Hamiltonian with a factorized state as its ground state. }
       \label{XYZphasehalf}
\end{figure}

{\it Half-integer spin: $s=1/2$ and $s=3/2$.}-  For $s=1/2$, there are four distinct phases, labeled as $\rm {AF}_x$, $\rm {AF}_y$, $\rm {AF}_z$, and $\rm {F}_z$, representing an antiferromagnetic phase in the $x$ direction, an antiferromagnetic phase in the $y$ direction, an antiferromagnetic phase in the $z$ direction, and a ferromagnetic phase in the $z$ direction, respectively.  We remark that the $\rm{AF}_\alpha$ phases, with $\alpha = x,y, z$,  are characterized in terms of a $Z_2$ symmetry-breaking order.  More precisely, these symmetry-breaking ordered phases are
characterized in terms of an order parameter: for the $\rm{F_z}$ phase, the order parameter is defined by $O_{F}^z=\langle S^z\rangle$; for the $\rm{AF_x}$ phase, the order parameter is defined by
$O_{AF}^x=\langle(-1)^i S^x_i\rangle$; for the $\rm{AF_y}$ phase, the order parameter is defined by
$O_{AF}^y=\langle(-1)^i S^y_i\rangle$;  for the $\rm{AF_z}$ phase, the order parameter is defined by
$O_{AF}^z=\langle(-1)^i S^z_i\rangle$.  Our numerical simulations yield the ground-state phase diagram, as shown in Fig.~\ref{XYZphasehalf}(a).  This is consistent with Baxter's exact solution~\cite{baxterbook}.
That is, for $s=1/2$, we are able to reproduce the ground-state phase diagram from the duality transformations, with the minimal knowledge of critical points in the principal regimes ${\rm I}$ and ${\rm {II}}$.

A remarkable fact is that, for $s=3/2$, the ground-state phase diagram, as plotted in Fig.~\ref{XYZphasehalf}(b), is identical to the ground-state phase diagram for $s=1/2$. Actually, this is valid for any half-integer spin $s$, since there is no other possibility to keep consistency with the Haldane conjecture~\cite{Haldane}.

\begin{figure}
     \includegraphics[angle=0,width=0.45\textwidth]{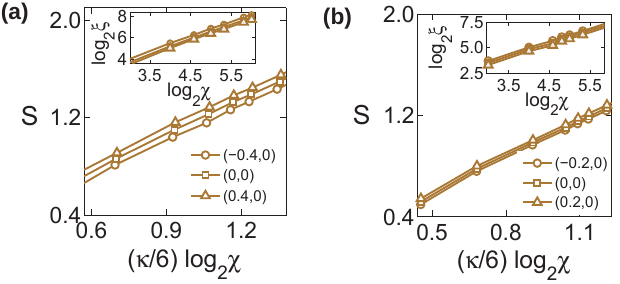}
        \caption{Scaling of the von Neumann entropy $S$ and the correlation length $\xi$ with respect to
the bond dimension $\chi$ for chosen critical points ($\Delta_c$, $\gamma_c$) in the quantum XYZ model: (a)  $s=1/2$; (b)  $s=3/2$.  Assuming $S(\chi) = c \kappa /6 \log_2{\chi}+a$, with $\kappa$ being determined from $\log_2{\xi} \sim \kappa\log_2(\chi)$ up to an additive constant (see insets), we may extract central charge $c$ for chosen critical points:
(a) $c= 1.0005, 0.9942$ and $0.9895$ for (-0.4, 0), (0, 0), (0.4, 0);
(b) $c= 0.9876, 0.9849$, and $0.9867$ for (-0.2, 0), (0, 0) and (0.2, 0).
Taking into account the numerical accuracies, we have $c=1$. } \label{centralchargehalf}
\end{figure}

There are five lines of critical points: $\gamma=0~(-1<\Delta\leq 1)$, $\gamma=1+\Delta~(\Delta<-1)$, $\gamma =   1 -\Delta ~(\Delta \geq 1)$, $\gamma=-1-\Delta~(\Delta<-1)$ and $\gamma =  - 1 +\Delta ~(\Delta \geq 1)$, which appear as the phase boundaries separating the  symmetry-breaking ordered phases $\rm {AF}_x$, $\rm {AF}_y$, $\rm {AF}_z$, and $\rm {F}_z$.  We have depicted three of them as the solid lines in Fig.~\ref{XYZphasehalf}, with the other two being symmetric under the mapping $\gamma \leftrightarrow -\gamma$.  A finite-entanglement scaling is performed, as shown in Fig~.\ref{centralchargehalf}, to extract central charge $c$ at a critical point (in the principal regimes ${\rm I}$ and ${\rm {II}}$). It is found that all the critical points are characterized in terms of central charge $c=1$, within the numerical accuracies.  Here, the bond dimension $\chi$ ranges from $8$ to $64$. For $s=1/2$,
this is consistent with the Bethe ansatz result that a line of critical points exists for $-1<\Delta\leq 1$ when $\gamma=0$, with central charge $c=1$.  If $\gamma =0$, a Kosterlitz-Thouless (KT) phase transition~\cite{kt} occurs at $\Delta=1$, protected by a $U(1)$ symmetry, from a critical regime to the $\rm {AF}_z$ phase for $\Delta>1$.  From the duality transformations and the mapping $\gamma \leftrightarrow -\gamma$, we see that a KT phase transition occurs at $\Delta=1$ along the lines $\gamma =-1+\Delta$ and $\gamma =1-\Delta$, again protected by a $U(1)$ symmetry.
Given central charge $c=1$, all the other phase transitions are Gaussian,  if the phase boundaries are crossed.

Although the five lines of critical points are self-dual under their respective duality transformations, the converse is not necessarily true.  Actually, the self-dual lines $\gamma=1-\Delta$ ($\Delta<1$), $\gamma=1+\Delta$ ($\Delta>-1$), $\gamma=-1-\Delta$ ($\Delta>-1$) and $\gamma=-1+\Delta$ ($\Delta<1$) represent {\it characteristic} lines that enjoy a $U(1)$ symmetry, in contrast to a $Z_2$ symmetry at a point away from these self-dual lines on the parameter space.
In addition, factorized states occur on the two self-dual lines $\gamma=1+\Delta$ ($\Delta>-1$) and $\gamma=-1-\Delta$ ($\Delta>-1$), as follows from an analysis of the geometric entanglement~\cite{ge}.  This is in agreement with a previous rigorous result about factorized states in the quantum XYZ model with arbitrary spin $s$~\cite{factorizedstates}.

{\it Integer spin: $s=1$ and $s=2$.}-  In addition to the four distinct symmetry-breaking ordered phases  $\rm {AF}_x$, $\rm {AF}_y$, $\rm {AF}_z$, and $\rm {F}_z$, an exotic phase-the Haldane phase - emerges surrounding the $SU(2)$ symmetric point $(\Delta =1, \gamma =0)$, as anticipated from the Haldane conjecture~\cite{Haldane}.   This phase may be characterized in terms of  a long-range string order parameter, defined by
\begin{equation}
O_s^{\alpha}=-\lim_{j-i \rightarrow \infty}\langle S^{\alpha}_i\exp{i\pi\sum_{i<k<j} S^{\alpha}_k}  S^{\alpha}_j\rangle,
\end{equation}
with $\alpha=x$, $y$ and $z$, respectively.

\begin{figure}
     \includegraphics[angle=0,width=0.225\textwidth]{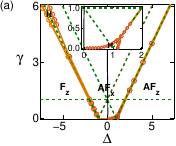}
\hspace{2mm}
     \includegraphics[angle=0,width=0.225\textwidth]{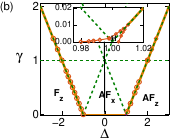}
       \caption{ Ground-state phase diagram for the quantum XYZ model with integer spin $s$: (a) $s=1$ and (b)  $s=2$. Here, the solid lines denote the phase boundaries between distinct symmetry-breaking ordered phases, $\rm {AF}_x$, $\rm {AF}_z$, $\rm {FM}_z$, and the Haldane phase.  The self-dual line $\gamma=1+\Delta$, with $\Delta>-1$, describes the Hamiltonian with a factorized state as its ground state.
      Inset (a): the Haldane phase for the quantum XYZ model with $s=1$ for $0\leq \gamma\leq 1$.
      Inset (b): the Haldane phase for the quantum XYZ model with $s=2$ for $0\leq \gamma<1$. Here, the bond dimension $\chi$ ranges from 200 up to 400.}
       \label{XYZphaseinteger}
\end{figure}

For $s=1$, our numerical simulations yield  the ground-state phase diagram, as plotted in Fig.~\ref{XYZphaseinteger}(a).
If $\gamma=0$, then the phase boundary between the Haldane phase and the $\rm{AF}_z$ phase is located at $\Delta_{c1}=1.185$, as determined from the iTEBD simulations, with the bond dimension $\chi=60$. This is in good agreement with a previous DMRG result: $\Delta_{c1}=1.186$~\cite{dmrgxxz}.  In addition, when $\gamma =0$, the phase boundary between the critical regime and the Haldane phase shifts towards $\Delta_{c2}=0.0$, as the bond dimension $\chi$  increases from 18 to 300. This  agrees with a previous DMRG result: $\Delta_{c2}=0$~\cite{dmrgxxz}.

For $s=2$,  the situation is similar.  However, the Haldane phase is shrinking with increasing $s$. This is sensible, given that the Haldane phase must vanish when the classical limit $s \rightarrow \infty$ is approached.  If $\gamma=0$, then the phase boundary between the Haldane phase and the $\rm{AF}_z$ phase is located at $\Delta_{c1}=1.0035$, as determined from the iTEBD simulations, with the bond dimension $\chi=200$. This is in good agreement with a previous DMRG result: $\Delta_{c1}=1.0037$~\cite{dmrgxxzspin2}.  In addition, when $\gamma =0$, the phase boundary between the critical regime and the Haldane phase is located at $\Delta_{c2}=0.98$, as determined from the bond dimension $400$. This is comparable to a previous DMRG result: $\Delta_{c2}=0.964$~\cite{dmrgxxzspin2}.

\begin{figure}
     \includegraphics[angle=0,width=0.44\textwidth]{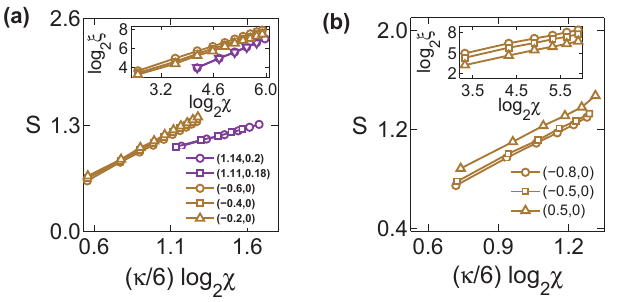}
       \caption{Scaling of the von Neumann entropy $S$ and the correlation length $\xi$ with respect to
the bond dimension $\chi$ for chosen critical points ($\Delta_c$, $\gamma_c$) in the quantum XYZ model: (a) $s=1$; (b)  $s=2$.
Assuming $S(\chi) = c \kappa /6 \log_2{\chi}+a$, with $\kappa$ being determined from $\log_2{\xi} \sim \kappa\log_2(\chi)$ up to an additive constant (see insets), we may extract central charge $c$ for chosen critical points:
 (a) $c= 0.515, 0.487, 1.0073, 1.0083$ and $1.0071$ for (1.14, 0.2), (1.11, 0.18), (-0.6, 0), (-0.4, 0) and (-0.2, 0);
 (b) $c= 0.9746, 0.9638$ and $0.9908$ for (-0.8, 0), (-0.5, 0) and (0.5, 0).
Taking into account the numerical accuracies, we have $c=1$ or $c=1/2$. } \label{centralchargeinteger}
\end{figure}

There are twelve lines of critical points, which appear as the phase boundaries separating the symmetry-breaking ordered phases, $\rm {AF}_x$, $\rm {AF}_y$, $\rm {AF}_z$, and $\rm {F}_z$, and the Haldane phase.  These lines of critical points fall into two distinct types: the first type consists of five lines of critical points separating the symmetry-breaking ordered phases: $\gamma=0~(-1<\Delta\leq \Delta_{c2})$, $\gamma=1+\Delta~(\Delta<-1)$, $\gamma =   1 -\Delta ~(\Delta \geq \Delta_{c1})$, $\gamma=-1-\Delta~(\Delta<-1)$ and $\gamma =  - 1 +\Delta ~(\Delta \geq \Delta_{c1})$,  and the second type consists of seven lines of critical points separating a symmetry-breaking ordered phase and the Haldane phase.
Note that only three lines of critical points  of the first type are depicted as the solid lines in Fig.~\ref{XYZphaseinteger}.
A finite-entanglement scaling is performed, as shown in Fig.~\ref{centralchargeinteger}, to extract central charge $c$ at a critical point (in the principal regimes ${\rm I}$ and ${\rm {II}}$). It is found that the five lines of critical points of the first type are characterized in terms of central charge $c=1$, and  the seven lines of critical points of the second type are characterized in terms of central charge $c=1/2$, within the numerical accuracies.  Here, the bond dimension $\chi$ ranges from $8$ to $64$.
When $\gamma =0$, a KT phase transition~\cite{kt} occurs at $\Delta_{c2}=0$ and $0.98$ for $s=1$ and $s=2$, respectively, protected by a $U(1)$ symmetry, from a critical regime to the Haldane phase.  From the duality transformations and the mapping $\gamma \leftrightarrow -\gamma$, we see that a KT phase transition occurs at $(2, 1)$ and $(2, -1)$ for $s=1$ and at $(1.01, 0.01)$ and $(1.01, -0.01)$ for $s=2$, respectively, along the lines $\gamma =-1+\Delta$ and $\gamma =1-\Delta$ , again protected by a $U(1)$ symmetry.
Given central charge $c=1$ or $c=1/2$, all the other phase transitions are either Gaussian or Ising-like,  if the phase boundaries are crossed.

As is well-known, in the Haldane phase,  there is a hidden $Z_2\times Z_2$ symmetry-breaking order~\cite{z2z2} for  odd-integer spin $s$,  but not for even-integer spin $s$~\cite{z2z2}. This involves a non-local unitary transformation with the $Z_2 \times Z_2$ symmetry.  Actually, for the transformed Hamiltonian, the phase corresponding to the Haldane phase is four-fold degenerate for $s=1$, but non-degenerate for $s=2$ under the $Z_2 \times Z_2$ symmetry group. This rules out the possibility that
there is any duality transformation between the Haldane phase and  the $\rm{AF}_\alpha$ phases, with $\alpha = x,y, z$.  This implies that the phase boundaries between the Haldane phase and  the $\rm{AF}_\alpha$ phases, with $\alpha = x,y, z$, are not self-dual under any duality transformation.

{\it Outlook.}- Our discussions about the duality transformations for the quantum XYZ model with arbitrary spin $s$ in one spatial dimension may be adapted to the quantum XYZ model with arbitrary spin $s$ on a bipartite lattice, such as the square and honeycomb lattices, as well as on a non-bipartite lattice, such as the triangular lattice, as  briefly discussed in Ref.~\cite{fm}. One may anticipate to map out the ground-state phase diagram for the quantum XYZ model on a specific lattice in the context of the infinite projected entangled-pair states (iPEPS)~\cite{ipeps}. This will be addressed in a forthcoming publication.

{\it Acknowledgement.}- We acknowledge enlightening discussions with Murray Batchelor, Vladimir Bazanov, Daniel Braak,  Gavin Brennen, Sam Young Cho, Clare Dunning, Jon Links, Vladimir Mangazeev, Javier Rodriguez-Laguna, and German Sierra.  This work was supported by Chongqing Research Program of Basic Research and Frontier Technology (Grant No. cstc2016jcyjA0480).


\begin{thebibliography}{10}


\bibitem{ortiz} H. Nishimori and G. Ortiz, Elements of Phase Transitions and Critical Phenomena (Oxford University Press, Oxford, 2011).

\bibitem{kw} H. A. Kramers and G. H. Wannier, Phys. Rev. \textbf {60}, 252 (1941);
J. B. Kogut, Rev. Mod. Phys. \textbf{51}, 659 (1979).

\bibitem{baxterbook} R.J. Baxter, Exactly Solved Models in Statistical Mechanics (Academic Press, 1982).
\bibitem{Haldane} F. D. M. Haldane, Phys. Lett. A \textbf{93}, 464  (1983); F. D. M. Haldane, Phys. Rev. Lett. \textbf{50}, 1153 (1983).
\bibitem{dmrgxxz}
 C. D. E. Boschi, E. Ercolessi, F. Ortolani, and M. Roncaglia, Eur. Phys. J. B \textbf{35}, 465 (2003);
 H. Ueda, H. Nakano, and K. Kusakabe, Phys. Rev. B \textbf{78}, 224402 (2008).
\bibitem{dmrgxxzspin2}
J. A. Kjall, M. P. Zaletel, R. S. K. Mong, J. H. Bardarson, and F. Pollmann, Phys. Rev. B \textbf{87}, 235106 (2013).

\bibitem{zhou} H.-Q. Zhou and J. P. Barjaktarevic, J. Phys. A: Math. Theor. \textbf{41}, 412001 (2008);
 H.-Q. Zhou, R. Or\'us, and G. Vidal, Phys. Rev. Lett. \textbf{100}, 080601 (2008);
 J.-H. Zhao and H.-Q. Zhou, Phys. Rev. B \textbf{80}, 014403 (2009).

\bibitem{white} S. R. White, Phys. Rev. Lett. \textbf{69}, 2863 (1992); U. Schollw\"{o}ck, Rev. Mod. Phys. \textbf{77}, 259 (2005).

\bibitem{vidal} G. Vidal, Phys. Rev. Lett. \textbf{91}, 147902 (2003); G. Vidal, Phys. Rev. Lett. \textbf{93}, 040502 (1-4) (2004); G. Vidal, Phys. Rev. Lett. \textbf{98}, 070201(1-4) (2007).
\bibitem{centralchargescaling}
L. Tagliacozzo, T. R. de Oliveira, S. Iblisdir, and J. I. Latorre, Phys. Rev. B \textbf{78}, 024410 (2008);
F. Pollmann, S. Mukerjee, A. M. Turner, and J. E. Moore, Phys. Rev. Lett. \textbf{102}, 255701 (2009).

\bibitem{ge} C.-Y. Huang and F.-L. Lin, Phys. Rev. A \textbf{81}, 032304 (2010);  Q.-Q. Shi, H.-L. Wang, S.-H. Li, S. Y. Cho, M. T. Batchelor, and H.-Q. Zhou, Phys. Rev. A \textbf{93},  062341 (2016).
\bibitem{kt} V. L. Berezinskii, Sov. Phys. JETP \textbf{34}, 610 (1971);
J. M. Kosterlitz and D. J. Thouless, J. Phys. C: Solid State Physics \textbf{6}, 1181 (1973).

\bibitem{factorizedstates}  S. M. Giampaolo, S. Montangero, F. Dell Anno, S. De Siena, and F. Illuminati, Phys. Rev. B \textbf{88}, 125142 (2013).

\bibitem{z2z2}  T. Kennedy and H. Tasaki, Phys. Rev. B \textbf{45}, 304 (1992); M. Oshikawa, J. Phys.: Condens. Matter \textbf{4} 7469 (1992).

\bibitem{fm} H.-Q. Zhou, Q.-Q. Shi and Y.-W. Dai, arXiv: 1709.09838.

\bibitem{ipeps} F. Verstraete, J. I. Cirac, and V. Murg, Adv.  Phys. \textbf{57}, 143 (2008);
J. I. Cirac and F. Verstraete, J. Phys. A: Math. Theor. \textbf{42}, 504004 (2009);
J. Jordan, R. Or\'us, G. Vidal, F. Verstraete, and J. I. Cirac, Phys. Rev. Lett. \textbf{101}, 250602 (1-4) (2008).
\end{thebibliography}
\end{document}